\documentclass[journal]{IEEEtran}

\usepackage[brazil,english]{babel}
\usepackage{amsthm,amsfonts,amsmath,amssymb}
\usepackage[ruled,vlined]{algorithm2e}
\usepackage{pifont}
\usepackage{subfigure} 
\usepackage{graphicx}
\usepackage{psfrag}

\usepackage{xcolor}
\usepackage{array}
\usepackage{units}
\usepackage{mdframed}
\usepackage{colortbl}
\usepackage{setspace}
\usepackage{textcomp}
\usepackage{multirow}
\usepackage{enumitem}
\usepackage{auto-pst-pdf}
\usepackage{booktabs} 
\usepackage{steinmetz}
\makeatletter
\newcommand{\mypm}{\mathbin{\mathpalette\@mypm\relax}}
\newcommand{\@mypm}[2]{\ooalign{%
  \raisebox{.1\height}{$#1+$}\cr
  \smash{\raisebox{-.6\height}{$#1-$}}\cr}}
\makeatother
\usepackage[ruled,vlined]{algorithm2e}
\begin{document}
%

\title{A Constraint Enforcement Deep Reinforcement Learning Framework  for Optimal Energy Storage Systems Dispatch}


%
%
%

\author{Shengren Hou,~\IEEEmembership{Student Member,~IEEE,}
        Edgar Mauricio Salazar Duque, {Student Member,~IEEE,}
        Peter~Palensky,~\IEEEmembership{Senior Member,~IEEE,}
        and  Pedro P. Vergara,~\IEEEmembership{Senior Member,~IEEE.}
\thanks{This work used the Dutch national e-infrastructure with the support of the
SURF Cooperative using grant no. EINF-4706.}
\thanks{Shengren Hou, Peter Palensky and Pedro P. Vergara are with the Intelligent Electrical Power Grids (IEPG) Section, Delft University of Technology, 2628 CD Delft, The Netherlands (email: \{h.hou-1, p.palensky, p.p.vergarabarrios\}@tudelft.nl)

Edgar Mauricio Salazar Duque is with the Electrical Energy Systems (EES) Group, Eindhoven University of Technology, 5612 AE Eindhoven, The Netherlands, (email: e.m.salazar.duque@tue.nl)}
}

%



\maketitle

\begin{abstract}
The optimal dispatch of energy storage systems (ESSs) presents formidable challenges due to the uncertainty introduced by fluctuations in dynamic prices, demand consumption, and renewable-based energy generation. By exploiting the generalization capabilities of deep neural networks (DNNs), deep reinforcement learning (DRL) algorithms can learn good-quality control models that adaptively respond to distribution networks' stochastic nature. However, current DRL algorithms lack the capabilities to enforce operational constraints strictly, often even providing unfeasible control actions. To address this issue, we propose a DRL framework that effectively handles continuous action spaces while strictly enforcing the environments and action space operational constraints during online operation. Firstly, the proposed framework trains an action-value function modeled using DNNs. Subsequently, this action-value function is formulated as a mixed-integer programming (MIP) formulation enabling the consideration of the environment's operational constraints. Comprehensive numerical simulations show the superior performance of the proposed MIP-DRL framework, effectively enforcing all constraints while delivering high-quality dispatch decisions when compared with state-of-the-art DRL algorithms and the optimal solution obtained with a perfect forecast of the stochastic variables. 
\end{abstract}


\IEEEpeerreviewmaketitle

\vspace{-3mm}
\section{Introduction}

\IEEEPARstart{T}{he} proliferation of distributed energy resources (DERs) poses various challenges in the control and operation of electrical distribution networks~\cite{aihui2022distributed}. For instance, overvoltage issues can be seen in networks with high photovoltaic (PV) penetration and low demand consumption (e.g., during noon). To overcome this, energy storage systems (ESSs) are increasingly being deployed, offering support services to the distribution system operators (DSOs) such as voltage magnitude regulation. These support services can be provided by exploiting ESSs' flexibility in response to a dynamic price throughout the day, which can be obtained by solving an optimal EESs scheduling problem. From the EESs operator's view, the defined EESs dispatch should minimize their operational cost, while respecting all voltage and current magnitude constraints of the distribution network. Nevertheless, such a scheduling problem is inherently challenging due to the stochastic and uncertain nature of the dynamic prices, the demand consumption, and the renewable-based generation (e.g.~from PV systems).

Traditional research in optimal ESSs dispatch has predominantly focused on developing accurate models and approximated formulations that make the problem amenable for commercial solvers (e.g.~\cite{Macedo2015}), collectively known as model-based approaches. For instance, mixed-integer nonlinear programming (MINLP) has been utilized to determine the optimal ESSs operation of an unbalanced microgrid~\cite{VergaraLopez2019}, linearizations have been introduced to reduce the complexity of the proposed formulations. In~\cite{chen2022robust}, a two-stage robust optimization framework has been proposed to simultaneously solve a day-ahead scheduling and real-time regulation problem of an integrated energy system. Nevertheless, these model-based approaches rely on the accuracy of their mathematical models and approximations, and their efficiency and performance often fall short, failing to provide high-quality solutions fast.

Model-free approaches have been proposed as an alternative to the shortcomings mentioned above. These approaches model the optimal EESs scheduling problem as a Markov Decision Process (MDP) and leverage reinforcement learning  (RL) algorithms to solve it~\cite{wang2021multi}. These RL algorithms generally deploy an agent that improves a control model based on a quality measure (reward) obtained by interacting with an environment~\cite{sutton_reinforcement_2018}. By exploiting the good generalization capabilities of deep neural networks (DNN), DRL algorithms can perform sequential interpretations of data, learning good-quality control models that can adaptively respond to the stochastic nature of an environment~\cite{mnih2015DQN}. Nevertheless, the optimal scheduling of ESSs requires meeting strict operational constraints, so safety and feasibility can be guaranteed, especially during online operations. The generalization nature of DRL algorithms, crucial for learning, means that DRL algorithms cannot directly enforce constraints on the environment and action space. 


Several approaches have been developed to improve the constraint enforcement capabilities of DRL algorithms~\cite{comprehensive_survey_SRL_review,benchmark_SRL_review}. For instance, to ensure that the updated policy stays within a feasible set, in \cite{cpo_overload_relief,achiam2017cpo}, a cumulative constraint violation index is kept below a predetermined threshold. This approach was also used in~\cite{cpo_distirbution_network}, in which the constraint violation index is designed to reflect the voltage and current magnitude violation level due to the ESSs dispatch defined. Nevertheless, enforcing constraints via cumulative indexes can only provide a probabilistic notion of safety, failing to enforce voltage and current magnitude constraints in real-time due to their instantaneous nature. Alternatively, a projection operator can be developed to project actions defined by the DRL algorithm into a feasible set~\cite{safe_ddpg}. For instance, the projection operator proposed in~\cite{kou2020safe,srl_ed} uses the action defined by the DRL algorithm as a starting point to solve a mathematical programming formulation, thereby ensuring constraints compliance. A similar approach was implemented in~\cite{kou2020safe} to regulate the distribution networks' voltage magnitude via the control of a smart transformer. Nevertheless, implementing such projection operators can degrade the DRL algorithm's performance, as shown in~\cite{srl_projection_performance_test}. Soft constraint enforcement is currently the most used approach for DRL algorithms~\cite{em_pd_DDPG}. In this approach, a large and fixed penalty term is incorporated in the reward function when training the parameters of the control policy. This allows the DRL algorithm to avoid actions that lead to unfeasible operations. For instance, in~\cite{pedro2022_rl_votlage_control}, the problem of dispatching PV inverters has been addressed by a decentralized framework that penalizes RL agents when their actions lead to voltage magnitude violations. Similarly, an RL algorithm with eligibility traces has been developed to dispatch ESSs systems, minimizing costs while regulating voltage magnitude~\cite{mauricio2022eligibility}. Although these strategies may enforce operational constraints during training, they cannot guarantee the feasibility of the defined operating schedule in real-time, especially during consumption and renewable-based generation peak periods~\cite{shengren2022performance}. 

As action feasibility, defined as the capability to enforce the environment's operational constraints, is paramount to ensure a wide implementation of DRL algorithms. In this paper, a constraint enforcement DRL framework named MIP-DRL is proposed. The proposed framework can strictly enforce all operational constraints, ensuring zero-constraint violations in real-time operation, even in unseen scenarios (data). This has been achieved by leveraging recent research optimization advancements for DNNs, which allows their representation as a mixed-integer linear (MIP) formulation. Thus, by utilizing the constraint-enforcing ability of mathematical programming, a theoretical guarantee for feasibility during online execution is provided. This paper extends our previous work in~\cite{shengren2023optimal}, evolving into a general framework that strictly enables state-of-the-art DRL algorithms to enforce operational constraints. The main contribution of this paper is the proposed framework itself, enabling DRL algorithms that effectively handle continuous action spaces to also enforce the environments and action space operational constraints strictly during online operation. As a result, the proposed framework enables different state-of-the-art DRL algorithms (i.e., DDPG, TD3, SAC) to ensure action feasibility even in unseen scenarios (data). In general, the proposed MIP-DRL framework is generic and can be extended to different state-of-the-art RL algorithms that use DNNs as a function approximation of a value-based function. Here, we developed versions of DDPG, TD3, and SAC algorithms capable of enforcing operational constraints strictly.

\vspace{-2mm}
\section{Mathematical Formulation of the Optimal EESs Scheduling Problem}\label{sec_nlp_formulation}
The optimal scheduling of ESSs in a distribution network can be modeled using the nonlinear programming (NLP) formulation given by \eqref{eq:goal}--\eqref{eq_SOC_cons}. The objective function in \eqref{eq:goal} aims to minimize the total operational cost over the time horizon~${\cal T}$, comprising the cost of importing power from the main grid. The operational cost $\rho_{t}$ at time slot $t$ is settled according to the day-ahead market prices $\rho_t$ in EUR/kWh.
\begin{equation}\label{eq:goal}
    \min_{\substack{P^{B}_{m,t}, \forall m \in {\cal B}, \forall t \in {\cal T}}} \left\lbrace  \sum_{t \in {\cal T}} \left[ \rho_{t}\sum_{m \in {\cal N}} \left(P^D_{m,t} + P^{B}_{m,t} - P^{PV}_{m,t}\right)\Delta t \right] \right\rbrace.
\end{equation}
Subject to:
%
\vspace{-2mm}
\begin{multline} \label{eq:active_power_balance}
 \hspace{-5mm} \sum_{nm \in {\cal L}} P_{nm,t} - \sum_{mn \in {\cal L}} (P_{mn,t} + R_{mn}I_{mn,t}^2) + P_{m,t}^{B} \\ + P_{m,t}^{PV}+ P_{m,t}^{S}= P_{m,t}^{D}  \quad \forall m \in {\cal N}, \forall t \in {\cal T}  
\end{multline}
\vspace{-6mm}
\begin{multline} \label{eq:reactive_power_balance}
 \hspace{-5mm} \sum_{nm \in {\cal L}} Q_{nm,t} - \sum_{mn \in {\cal L}} (Q_{mn,t} + X_{mn}I_{mn,t}^2) + Q_{m,t}^{S} = Q_{m,t}^{D}  \\  \forall m \in {\cal N}, \forall t \in {\cal T}
\end{multline}
\vspace{-6mm}
\begin{multline}
\label{eq_votlage_drop}
 \hspace{-4mm} V_{m,t}^2-V_{n,t}^2=2(R_{mn}P_{mn,t}+X_{mn}Q_{mn,t})+ \\(R_{mn}^2+X_{mn}^2)I_{mn,t}^2 \quad  \forall m,n \in {\cal N}, \forall t \in {\cal T}  
\end{multline}
\vspace{-6mm}
\begin{flalign}
& V_{m,t}^2I_{mn,t}^2=P_{mn,t}^2+Q_{mn,t}^2 & \forall m,n \in {\cal N}, \forall t \in {\cal T} \label{eq_vi=pq} &     
\end{flalign}
\vspace{-6mm}
\begin{multline}
\hspace{-4mm} SOC_{m,t}^{B}=SOC_{m,t-1}^{B} + \eta^{B}_{m}P_{m,t}^{B}\Delta t/\overline{E}^{B}_{m} \hspace{0.30em} \forall m \in {\cal{B}}, \forall t \in {\cal{T}} \hspace{-0.4em} \label{eq_SOC_cha}
\end{multline}
\vspace{-6mm}
\begin{flalign}
& \underline{SOC}_{m}^{B}\leq SOC_{m,t}^{B}\leq\overline{SOC}_{m}^{B} & \forall m \in {\cal{B}}, \forall t \in {\cal{T}} & \label{eq_SOC_cons}\\
& \underline{P}^{B}_{m}\leq P^{B}_{m,t}\leq \overline{P}^{B}_{m} & \forall m \in {\cal B}, \forall t \in {\cal T} & \label{eq_char_disc_cons}\\
& \underline{V}^{2}\leq V_{m,t}^2\leq \overline{V}^{2} & \forall m \in {\cal N}, \forall t \in {\cal T} & \label{eq:voltage_boundary}\\
& 0 \leq I_{mn,t}^2 \leq \overline{I}_{mn}^{2} & \forall mn \in {\cal L}, \forall t \in {\cal T} & \label{eq:limites_corre_5}\\
& P_{m,t}^{S} = Q_{m,t}^{S} = 0 & \forall m \in {\cal N} \backslash \{1\}, \forall t \in {\cal T} & \label{eq:power_not_substation}
\end{flalign}

The distribution network is modeled using the power flow formulation shown in \eqref{eq:active_power_balance}--\eqref{eq_vi=pq} in terms of the active $P_{mn,t}$ power, reactive power $Q_{mn,t}$ and current magnitude $I_{mn,t}$ of lines, and the voltage magnitude $V_{m,t}$ of nodes. Equation in \eqref{eq_SOC_cha} models the dynamics of the ESSs' SOC on the set ${\cal B}$, while \eqref{eq_SOC_cons} enforces the SOC limits. Hereafter, it is assumed that the ESS $m \in {\cal B}$ is connected to node $m$, thus, ${\cal B} \subseteq {\cal N}$. Finally, \eqref{eq_char_disc_cons} enforces the  ESSs discharge/charge operation limits, \eqref{eq:voltage_boundary} and \eqref{eq:limites_corre_5} enforce the voltage magnitude and line current limits, respectively, while \eqref{eq:power_not_substation} enforces that only one node is connected to the substation. Notice that to solve the above-presented NLP formulation, all long-term operational data (e.g., expected PV generation and consumption) must be collected in order to properly define the EESs' dispatch decisions, while the power flow formulation must also be considered in order to enforce the voltage and current magnitude limits.

\section{ESSs Scheduling Problem MDP Formulation}\label{sec:mdp_formulation}

The above-presented mathematical formulation can be modeled as a finite MDP, represented by a 5-tuple $(\cal{S},\cal{A},\cal{P},\cal{R},\gamma)$, where $\cal{S}$ represents a state space, $\cal{A}$ an action space, $\cal{P}$ the state transition probability function, $\cal{R}$ the reward function, and $\gamma$ a discount factor. The decision as to which action $a_t$ is chosen in a particular state $s_t$ is governed by a policy $\pi(a_t|s_t)$. In a standard RL algorithm, an RL agent employs the policy $\pi(a_t|s_t)$ to interact with the formulated MDP defining a trajectory of states, actions, and rewards: $\tau=\left(\boldsymbol{s}_0, \boldsymbol{a}_0, \boldsymbol{s}_1, \boldsymbol{a}_1, \cdots\right)$. Here, the RL agent's goal is to estimate a policy that maximizes the cumulative discounted return $J(\pi)=\mathbb{E}_{\tau \sim \pi}\left[\sum_{t=0}^{{\cal T}} \gamma^t r_t\right]$, in which ${\cal T}$ is the length of the control horizon. 

Different from the standard RL approach, in a constrained MDP, the RL agent aims to estimate a policy $\pi$ confined in a feasible set $\Pi_{C}=\left\{\pi: J_{C_{i}}(\pi) \leq 0, \quad i=1, \ldots, k\right\}$. Here, $J_{C_{i}}(\pi)$ denotes a cost-based constraint function induced by the constraint violation functions $C_{i,t}(\cdot),~i=1,\cdots,k$. Based on these definitions,  a constrained MDP can be formulated as the next constrained optimization problem:
\begin{equation}\label{eq:constrained_MDP}
\begin{aligned}
&\max _{\pi} J(\pi)=\mathbb{E}_{\tau \sim \pi}\left[\sum_{t=0}^{{\cal T}} \gamma^t r_t\right] \\
&\text { s.t. } J_{C_i}(\pi) \leq 0, \forall i=1, \ldots, k. \\
\end{aligned}
\end{equation}
\noindent Here, $J_{C_i}(\pi)$ is defined as $J_{C_i}(\pi)=\mathbb{E}_{\tau \sim \pi}\left[\sum_{t=0}^{{\cal T}} \gamma^t C_{i,t}\right] $. A more detailed MDP description of the EESs optimal scheduling problem is presented below.

\subsection{State}
The state $s_t$ denotes the operating status information of the network, which the agent can observe. The state at $t$ is defined by $s_t=[P^{N}_{m,t}|_{m \in {\cal N}},\rho_t,SOC^{B}_{m,t}|_{m \in {\cal B}}]$, where $P^{N}_{m,t}=P^{D}_{m,t}-P^{PV}_{m,t}$ corresponds to the nodal net power. These features can be divided into endogenous and exogenous features. The former includes the PV generation $P^{PV}_{m,t}$ and consumption $P^{D}_{m,t}$, and day-ahead price $\rho_t$, which are independent of the operated actions, while the latter includes the ESSs' SOC $SOC^{B}_{m,t}$, which depends on the agent's action. 

\subsection{Action}
The action at time $t$ is defined as $a_t=[P^{B}_{m,t}|_{m \in {\cal B}}$], in which $a_t\in {\cal A}$, while ${\cal A}$ is a continuous space. Notice that $a_t$ refers to the charging/discharging dispatch for the $m_{th}$ ESS connected to node $m$ in the distribution network. 

\subsection{State Transition Function}
Given the state $s_{t}$ and action $a_{t}$ at time step $t$, the system transit to the next state $s_{t+1}$ defined by the next transition probability 
\vspace{-2mm}
\begin{multline} \label{eq:transition_function}
  p(S_{t+1},R_{t}|S_t,A_t)= \\ 
    \operatorname{Pr}\left\{S_{t+1}=s_{t+1}, R_{t}=r_{t} \mid S_{t}=s_{t}, A_{t}=a_t\right\}.  
\end{multline}

\noindent This transition probability function $\cal{P}$ models the endogenous distribution network dynamics, determined by the electrical network itself, and the exogenous uncertainty caused by the PV generation, demand consumption, and the day-ahead price dynamics. In practice, building an accurate mathematical model for such a transition function is not possible. Nevertheless, model-free RL algorithms do not require prior knowledge of  function $\cal{P}$ as it can be learned by interacting with the environment. 

\subsection{Reward Function} 
RL algorithms can learn representative operation strategies from interactions with the environment. To achieve this goal, the environment must provide a reward $r_{t}$ to quantify the goodness of any action taken during the interaction process. In this case, the raw reward is defined as the negative of the operational cost for the operation of the distribution network, modeled defined as: 
\begin{equation}\label{eq_reward}
\hspace{-2mm} {\cal R}_{t}\left(s_{t}, a_{t}\right)= r_t =  - \rho_{t} \left[ \sum_{m \in {\cal N}} \left( P^D_{m,t}+P^{B}_{m,t}+P^{PV}_{m,t}\right)  \right] \Delta t
\end{equation}


\subsection{Operational Constraints}
The developed RL algorithm minimizes the total operational cost while enforcing all the ESSs and distribution network's operational constraints. Thus, the defined ESSs dispatch must respect the following set of operational constraints: first, the maximum discharge/charge capacity, as is expressed in~\eqref{eq_char_disc_cons}; second, constraints in~\eqref{eq_SOC_cons} require the EESs SOC to operate within the predefined SOC range; and third, the voltage and current magnitude across the electrical distribution network should also be within its feasible values, as modeled in~\eqref{eq:voltage_boundary} and \eqref{eq:limites_corre_5}, respectively.

Operational constraints in~\eqref{eq_SOC_cons} and~\eqref{eq_char_disc_cons} can be directly met by enforcing the RL algorithms' boundaries of delivered actions $a_t$. This corresponds to the standard procedure to enforce such constraints on the action space ${\cal A}$, as shown in~\cite{shengren2022performance}. Nevertheless, constraints in~\eqref{eq:voltage_boundary} and \eqref{eq:limites_corre_5} involve the physical dynamics of the distribution network, which RL algorithms can not directly handle. Notice that \eqref{eq:constrained_MDP} does not specify how the set of constraints $C_{i},~i=\{1,\cdots,k\}$ can be enforced. To enforce constraints in~\eqref{eq:voltage_boundary} and \eqref{eq:limites_corre_5} in the MDP formulation, the constraint violation functions $C_{m,t}$ are added as penalty terms to the raw reward function in~\eqref{eq_reward}. This procedure transforms the constrained optimization problem in \eqref{eq:constrained_MDP} into an unconstrained one, further reformulated as:
\begin{multline}\label{eq:new_penaly_reward}
r_t =  - \rho_{t} \left[ \sum_{m \in {\cal N}} \left( P^D_{m,t}+P^{B}_{m,t}-P^{PV}_{m,t}\right) \right] \Delta t \\ -\sigma \left[  \sum_{m\in\cal{B}} C_{m,t}(V_{m,t})\right] , 
\end{multline}

\noindent and in which $\sigma$ is used to define the trade-off between the operational cost and the penalty incurred by voltage magnitude violations. The constraint violation function $C_{m,t}$ in \eqref{eq:new_penaly_reward} can be modeled using different functions (e.g., $L_2$ functions). Here, as in~\cite{mauricio2022eligibility}, $C_{m,t}$ is defined as
\begin{equation}
    C_{m,t}=\min \left\{0, \left(\frac{\overline{V}-\underline{V}}{2}-\left|V_0-V_{m, t}\right|\right)\right\}, \forall m \in {\cal B}.
\end{equation}

\noindent It is critical to notice that enforcing operational constraints by only adding a penalty term into the reward function can lead to infeasible operational states, as observed in~\cite{mauricio2022eligibility}. This is since the constrained MDP formulation in \eqref{eq:constrained_MDP} does not account for requirements to satisfy operational constraints after training. To overcome this, we introduce a constraint enforcement framework for DRL algorithms. We denominate such a framework as MIP-DRL, as it exploits MIP theory for deep learning models. In this framework, by utilizing the constraint-enforcing ability of mathematical programming, a theoretical guarantee for feasibility during online execution is provided.

\section{Constraint Enforcement MIP-DRL Framework}

The proposed MIP-DRL framework is defined through two main procedures: \textit{(i) Training}, where the action-value function is approximated, and \textit{(ii) Deployment}, executed during online decision-making. Both of these procedures are explained in detail below. 



\subsection{Training Procedure}\label{value_iteration_procedure}

\subsubsection{Value-Based RL Algorithms}
Define $Q_{\pi}(S_t,A_t)$ as the action-value function that estimates the expected accumulated reward given that action $a_t$ is taken at state $s_t$ and following policy $\pi(\cdot)$ after that. The action-value function $Q_{\pi}(S_t,A_t)$ can be expressed recursively as~\cite{sutton_reinforcement_2018},
\begin{multline} \label{eq_recursive_q}
Q_{\pi}(S_t,A_t)= \mathbb{E}_{\pi} \left[r_t + \right. \\ \left. \gamma Q_{\pi}(s_{t+1},a_{t+1})|S_t=s_t,A_t=a_t\right]  
\end{multline}

Bellman's principle of optimality states that the optimal action-value function for an MDP has the recursive expression
\begin{multline}\label{eq_recursive_q_2}
    Q_{\pi}^{*}(S_t,A_t) =  \mathbb{E}_{\pi}[r_t  \\ + \gamma \max_{\substack{a_{t+1} \in {\cal A}}} Q_{\pi}^{*}(s_{t+1},a_{t+1})|S_t=s_t,A_t=a_t],
\end{multline}

\noindent which solution can be obtained by using a Temporal Difference (TD) algorithm~\cite{qlearning_watkins}, which iteratively solves the following update rule.
\begin{multline}\label{eq_recursive_q_approx}
    \hat{Q}(S_t,A_t) \doteq \hat{Q}(S_t,A_t)  + \\ \alpha \left[r_{t} + \gamma \max_{\substack{a_{t+1} \in {\cal A}}} \hat{Q}(s_{t+1},a_{t+1}) - \hat{Q}(S_t,A_t)\right],
\end{multline}
in which $\hat{Q}(\cdot)$ corresponds to a function approximator used to represent $Q_{\pi}^{*}(\cdot)$ and $\alpha \in (0,1]$ is a learning rate. Once a good quality representation of $Q_{\pi}^{*}(\cdot)$ is obtained via $\hat{Q}(\cdot)$, at time step $t$ and state $s_t$, optimal actions $a_{t}$ can be sampled from the optimal policy, i.e., $a_t \sim \pi^{*}(s_t) $, obtained as 
\begin{equation}\label{eq_optimal_action}
\pi^{*}(S_t) = \max_{\substack{a \in {\cal A}}} \hat{Q}(S_t=s_t,a).
\end{equation}

\noindent For continuous state and action spaces, the optimal action-value function $Q_{\pi}^{*}(\cdot)$ can be approximated using a DNN i.e., $\hat{Q}(\cdot)=Q_{\theta}(\cdot)$ with parameters $\theta$. This corresponds to the training procedure's main objective, which can now be seen as a regression problem aiming to estimate the DNN's parameters $\theta$. However, notice that in continuous action spaces, the procedure used in \eqref{eq_optimal_action} to sample actions from the action-value function $Q_{\theta}$ is not feasible since a directly exhaustive enumeration from DNNs (i.e., the $max-Q$ problem) is complex. A DNN policy $\pi_{\omega}(\cdot)$ with parameters ${\omega}$ is leveraged to overcome this, leading to DRL algorithms known as \textit{actor-critic} algorithms. How these actor and critic models are updated and interact with the environment is explained next. 

\subsubsection{Step-by-Step Training}
\begin{figure*}[ht]
   \centering
    \psfrag{t1}[][][0.7]{$\text{Q-function}$}
    \psfrag{t2}[][][0.7]{$\text{Policy}$}
    \psfrag{t3}[][][0.8]{$\omega$}
    \psfrag{a1}[][][0.8]{$Q_{\theta}(\cdot)$}
    \psfrag{a2}[][][0.8]{$\pi_{\omega}(\cdot)$}
    \psfrag{a3}[][][0.8]{$\textbf{Replay}$}
    \psfrag{a4}[][][0.8]{$\textbf{Buffer}$}
    \psfrag{a5}[][][0.8]{$\textbf{Interaction}$}
    \psfrag{a6}[][][0.5]{$\textbf{Environment}$}
    \psfrag{a7}[][][0.8]{$(a)$}
    \psfrag{b1}[][][0.8]{$\theta$}
    \psfrag{b2}[][][0.8]{$\nabla_{\theta}$}
    \psfrag{b3}[][][0.8]{$\nabla_{\omega}$}
    \psfrag{b4}[][][0.7]{$[s_t,a_t,r_t,s_{t+1}]$}
    \psfrag{b5}[][][0.8]{$s_{t+1}$}
    \psfrag{b6}[][][0.8]{$r_t$}
    \psfrag{b7}[][][0.8]{$s_{t}$}
    \psfrag{b8}[][][0.8]{$a_t$}
    \psfrag{b9}[][][0.7]{$a_t\sim\pi_{\omega}(s)$}
    \psfrag{b10}[][][0.8]{$\textbf{Environment}$}
    \psfrag{b12}[][][0.7]{$\text{State Transition}$}
    \psfrag{b13}[][][0.7]{$\text{System}$}
    \psfrag{b14}[][][0.7]{$\text{Measurements}$}
    \psfrag{b15}[][][0.7]{$\text{System}$}
    \psfrag{b16}[][][0.7]{$\text{Parameters}$}
    \psfrag{b17}[][][0.7]{$\text{Battery}$}
    \psfrag{b18}[][][0.7]{$\text{Dispatch}$}
    \psfrag{b19}[][][0.7]{$\text{Power Flow}$}
    \psfrag{b20}[][][0.7]{$\text{Solver}$}
    \psfrag{b21}[][][0.7]{$\text{Reward}$}
    \psfrag{b22}[][][0.7]{$(b)$}
    \psfrag{b23}[][][0.7]{$\text{Explore policy}$}
    \psfrag{b24}[][][0.7]{$\text{Q-function}$}
    \psfrag{b25}[][][0.7]{$(c)$}
    \psfrag{b26}[][][0.8]{$s_{t+1}$}
    \psfrag{M1}[][][0.8]{$ \pi_\omega(s)$}
    \psfrag{Interaction}[][][0.8]{$\text{Interaction}$}  
    \psfrag{M2}[][][0.8]{$Q_\theta(s,a)$}
\includegraphics[width=1.8\columnwidth]{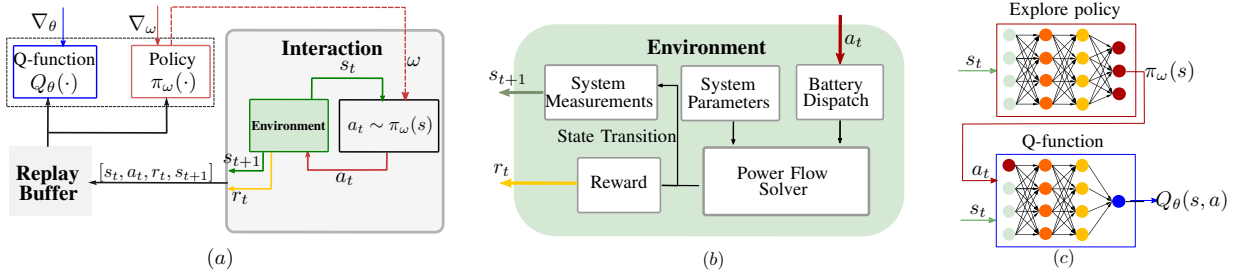}
   \caption{Training procedure of the proposed MIP-DRL framework: $(a)$ Interaction with the environment is done by sampling actions from a policy model $\pi_{\omega}(s)$. Information collected from the environment in the form of tuples $(s_t,a_t,r_t,s_{t+1})$ are stored in a replay buffer $R$ and later used to update the parameters of the policy $\pi_{\omega}(s)$ (or actor) and action-value function $Q_{\theta}(s,a)$ (or critic) model. $(b)$ Environment (distribution network) composed of a power flow solver. After implementing the current actions $a_t$ (ESSs dispatch schedule), the environment provides the reward $r_t$ as in \eqref{eq:new_penaly_reward} and the state transition to define $s_{t+1}$ via (simulated) network measurements. $(c)$ Policy model $\pi_{\omega}(s)$, which defines the action $a_t$, and the action-value function $Q_{\theta}(s,a)$, which assess the quality of the defined action $a_t$ for state $s_t$.}
   \label{fig_Q_iteration_framework}
\end{figure*}

Algorithm~\ref{algorithm1} describes the step-by-step procedure implemented, while Fig.~\ref{fig_Q_iteration_framework} illustrates the interaction of the actor $\pi_{\omega}(\cdot)$ (also known as policy) and critic $Q_{\theta}(\cdot)$ (also known as action-value function) models with the environment (distribution network). To start the training procedure, the parameters of the DNN for the critic~$Q_{\theta}(\cdot)$ and the actor~$\pi_{\omega}(\cdot)$ models are randomly initialized. Then, as shown in Fig.~\ref{fig_Q_iteration_framework}$(a)$, interaction with the environment is performed by sampling an action $a_t$ (with an exploration noise) from the actor model $\pi_{\omega}(\cdot)$ for the state $s_t$, causing the environment to move to the next state $s_{t+1}$, and observing a reward $r_t$ as a measure of the quality of the action $a_t$ implemented, see Fig.~\ref{fig_Q_iteration_framework}$(b)$. This procedure is done for a predefined number of training epochs $L$. During these interactions, the transition tuples $(s_t,a_t,r_t,s_{t+1})$ are stored in a replay buffer $R$. Later, a subset $B$ of these samples is selected and used to update the parameters of the policy $\pi_{\omega}(\cdot)$ (actor) and the value function $Q_{\theta}(\cdot)$ (critic) models. This corresponds to the next step, explained next. 

\begin{algorithm}[t]\label{algorithm1}
	\caption{Training procedure for the proposed MIP-DRL framework}
	Define the maximum training epochs $L$, and episode length ${\cal T}$. 
	Initialize parameters of DNNs $Q_{\theta}$, $Q_{\theta^{\text{target}}}$, and $\pi_\omega$; 
	Initialize reply buffer $R$ \;
    \While{Training}{\For {$l=1$ \KwTo $L$}{
		Sample an action with exploration noise $a_t \sim \pi_\omega(s_t)+\epsilon$,
		$\epsilon \sim \mathcal{N}(0, \sigma)$ \; 
            Observe reward $r_t$ and new state $s_{t+1}$ \;
	    \par
	    Store transition tuple $\left(s_t, a_t, r_t, s_{t+1}\right)$ in $R$\;}}

     \textbf{Policy and Action-Value Function Update}\;
          
    \While{Parameters Update}{
    \For {$t=1$ \KwTo $T$}{Sample a random mini-batch of $|B|$ transitions $\left(s_t, a_t, r_t, s_{t+1}\right)$ from $R$\;
		\par
  Execute Algorithm~\ref{MIP-DDPG} or Algorithm~\ref{MIP-TD3} or Algorithm~\ref{MIP-SAC}\;} 
    } 
\end{algorithm}

\subsubsection{Policy and Action-Value Function Update}
Different DRL algorithms are defined depending on how the policy $\pi_{\omega}(s)$ and the action-value function $Q_{\theta}(s,a)$ network interact and are updated~\cite{actor_expert}. Thus, the proposed MIP-DRL framework defines Algorithm~\ref{MIP-DDPG} for the MIP-DDPG algorithm, based on the DDPG algorithm in~\cite{lillicrap_ddpg}; Algorithm~\ref{MIP-TD3} for the MIP-TD3 algorithm, based on the TD3 algorithm in~\cite{fujimoto_td3_2018}; and Algorithm~\ref{MIP-SAC} for the MIP-SAC algorithm, based on the SAC algorithm in~\cite{haarnoja2018softactor}. These state-of-the-art DRL algorithms are used as they use DNNs as function approximations of an action-value function.  

Algorithm~\ref{MIP-DDPG} shows the updating steps followed by the MIP-DDPG algorithm, which samples an action $a_t$ from the actor model based on the current state $s_t$, while the critic model estimates the expected total reward for the current state-action pair $(s_t,a_t)$. The actor model's parameters are updated to maximize the critic's output, which represents the expected total reward as defined in~\eqref{eq_DDPG_policy}, 
\begin{equation}\label{eq_DDPG_policy}
\omega\leftarrow \omega+\nabla_{\omega} \frac{1}{|B|} \sum_{s \in B} Q_{\theta}\left(s, \pi_{\omega}(s)\right)   
\end{equation}

\noindent while the critic model's parameters are updated using a TD error, as defined in~\eqref{eq_DDPG_Q_update}, where $Q_{\theta^{\text{target}}} $ is a \textit{target} Q-function\footnote{i.e., a copy of model $Q_{\theta}$ which parameters are updated less frequently. This procedure helps to stabilize learning within the DRL algorithm. For a more detailed explanation, see~\cite{lillicrap_ddpg}.}. 
\begin{equation}
    \label{eq_DDPG_Q_update}
    \min _{\theta} \sum_{s \in B}\left(r_{t}+\gamma Q_{\theta^{\text {target}}}\left(s_{t+1}, \pi_{\omega}(s_{t+1})\right) - Q_{\theta}\left(s_{t}, a_{t}\right)\right)^{2}
\end{equation}

Algorithm~\ref{MIP-TD3} shows the updating procedure followed by the MIP-TD3 algorithm. A separate target network and double Q-networks are used to reduce the overestimation bias. In this case, the updates of the actor model are done by using
\begin{equation}\label{eq_TD3_policy}
\omega\leftarrow \omega+\nabla_{\omega} \frac{1}{|B|} \sum_{s \in B}
\Biggl( \min_{i=1,2} \{ Q_{\theta_i}\left(s, \pi_{\omega}(s)\right) \} \Biggr)
\end{equation}

\noindent while the critic update is done by using
\begin{equation}
    \label{eq_TD3_Q_update}
    \min _{\theta} \sum_{s \in B}\left(r_{t}+\gamma \min_{i=1,2} \{Q_{\theta_{i}^{\text {target}}}\left(s_{t+1}, \pi_{\omega}(s_{t+1})\right) \} - Q_{\theta_{i}}\left(s_{t}, a_{t}\right)\right)^{2}
\end{equation}

\noindent Finally, Algorithm~\ref{MIP-SAC} defines the updating steps for the MIP-SAC algorithm, which uses a soft Q-function to estimate the expected return, defined as in~\eqref{eq_SAC_Q_update}. 
\begin{multline}\label{eq_SAC_Q_update}
  \min _{\theta} \sum_{s \in B} \Biggl(r_t + \min_{i=1,2} \{ Q_{\theta_{i}^{\text{target}}}\left(s_{t+1}, \pi_{\omega}(s_{t+1})\right) \} \\
- \alpha \log \pi_\omega\left(a_{t} \mid s_{t}\right) - Q_{\theta_i}\left(s_{t}, a_{t}\right)\Biggr)^{2}
\end{multline}

\noindent Then, the policy model is updated based on minimizing the KL divergence between the policy and a target distribution, defined as in~\eqref{eq_SAC_policy}.   
\begin{multline}\label{eq_SAC_policy}
\omega\leftarrow \omega+\nabla_{\omega}\frac{1}{|B|} \sum_{s \in B} \Biggl(
\min_{i=1,2}\{ Q_{\theta_i}\left(s_t, \pi_{\omega}(s_t)\right)\} \\
-\alpha \log \pi_\omega\left(a_t \mid s_{t }\right) \Biggr)
\end{multline}

\vspace{-3mm}
\begin{algorithm}[ht]\label{MIP-DDPG}
\caption{Update MIP-DDPG Algorithm}
Action selection: $a_t \sim \pi_\omega(s_t)+\epsilon$,
		$\epsilon \sim \mathcal{N}(0, \sigma)$\;
  Update the Q-function Parameters~\eqref{eq_DDPG_Q_update}\;
  Update the policy function parameters by~\eqref{eq_DDPG_policy} \;
  Update the target-Q function parameters using~\eqref{eq_DDPG_Q_update}
\end{algorithm}

\vspace{-3mm}
\begin{algorithm}[ht]\label{MIP-TD3}
\caption{Update MIP-TD3 Algorithm}
Action selection: $a_t \sim clip \{\pi_\omega(s_t)+clip\{\epsilon, -c, c\}, a_{L}, a_{H}\}$,
		$\epsilon \sim \mathcal{N}(0, \sigma)$\;
  Update the Q-function Parameters:~\eqref{eq_TD3_Q_update}\;
  Update the policy function parameters by~\eqref{eq_TD3_policy}\;
  Update the target-Q function parameters using~\eqref{eq_TD3_Q_update}
\end{algorithm}

\vspace{-3mm}
\begin{algorithm}[ht]\label{MIP-SAC}
\caption{Update MIP-SAC Algorithm}
Action selection: $a_t \sim \pi_\omega(s_t)$\;
  Update the Q-function parameters~\eqref{eq_SAC_Q_update}\;
  Update the policy function parameters by~\eqref{eq_SAC_policy}\;
  Update the target-Q function parameters using~\eqref{eq_SAC_Q_update}
\end{algorithm}

\subsection{Deployment Procedure}

Once a good quality policy $\pi_{\omega}(\cdot)$ and action-value function $Q_{\theta}(\cdot)$ models are obtained, the \textit{Deployment} procedure can be executed. In this procedure phase, the policy model $\pi_\omega(\cdot)$ is discarded, and only the action function model $Q_{\theta}(\cdot)$ is used. Nevertheless, actions defined using only such action function $Q_{\theta}(\cdot)$ cannot strictly enforce operational constraints. To overcome this, we exploit recent research optimization advancements for DNNs. Thus, proposing a transformation of the DNN  $Q_{\theta}(\cdot)$  model into a MIP formulation.

\subsubsection{MIP representation of DNNs}\label{sec:explain_DNN}
Consider a DNN with $K+1$ layers listed from 0 to $K$. Layer 0 represents the DNN's input, while the last layer, $K$, represents the output. Each layer $k\in \{0,1,\dots,K\}$ has $U_k$ units, denoted by $u_{j,k}$, with $j$ being the unit index in layer $k$. We denote the output vector of layer $k$ as $x^k$, where $x^k_{j}$ is the output of unit $u_{j,k},~(j=1,2,\dots,U_k)$. Since layer 0 is the input of the DNN, $x_j^0$ represents the $j_{th}$ input value for the DNN. For each layer $k\leq1$, the unit $u_{j,k}$ computes the output vector $x^k$ as follows:

\begin{equation}\label{eq_general_x_output}
x^{k}=h\left(W^{k-1} x^{k-1}+b^{k-1}\right).
\end{equation}

\noindent Here, $W^{k-1}$ and $b^{k-1}$ are matrices of weights and biases that make up the set of parameters $\theta$, denoted as $\theta=\{W,b\}$. $h(\cdot)$ is the activation function corresponding to the ReLU function. The ReLU function is defined as follows: for a real vector $y$, $\operatorname{ReLU}(y):=\max \{0, y\}$.

Based on these definitions, the value function DNN $Q_{\theta}(\cdot)$ obtained from the proposed MIL-DRL algorithms, and with fixed parameters $\theta$, can be modeled as a valid MIP problem by modeling the ReLU function using binary constraints. Thus, using a binary activation variable $z^k_j$ for each unit $u_{j,k}$, the DNN $Q_{\theta}(\cdot)$ can be expressed as~\cite{fischettiJo2018}, 

\begin{equation}\label{eq_goal}
\min_{\substack{x_j^k, s_j^k, z_j^k, \forall k}}  \left\{ \sum_{k=0}^{K} \sum_{j=1}^{l_{k}} c_{j}^{k} x_{j}^{k}+\sum_{k=1}^{K} \sum_{j=1}^{l_{k}} d_{j}^{k} z_{j}^{k} \right\}
\end{equation}
Subject to:
\begin{equation}\label{eq_relu_milp}
    \left.\begin{array}{r}
    \sum_{i=1}^{l_{k-1}} w_{i j}^{k-1} x_{i}^{k-1}+b_{j}^{k-1}=x_{j}^{k}-s_{j}^{k} \\
    x_{j}^{k}, s_{j}^{k} \geq 0 \\
    z_{j}^{k} \in\{0,1\} \\
    z_{j}^{k}=1 \rightarrow x_{j}^{k} \leq 0 \\
    z_{j}^{k}=0 \rightarrow s_{j}^{k} \leq 0
    \end{array}\right\} \\
    \forall k, \forall j,
\end{equation}

\vspace{-2mm}
\begin{equation}
\label{eq_bound_l0}
    l b_{j}^{0} \leq x_{j}^{0} \leq u b_{j}^{0}, \quad j \in l_{0},
\end{equation}

\vspace{-4mm}
\begin{equation}\label{eq_bound_k}
    \left.\begin{array}{l}
    l b_{j}^{k} \leq x_{j}^{k} \leq u b_{j}^{k} \\
    \overline{l b}_{j}^{k} \leq s_{j}^{k} \leq \overline{u b}_{j}^{k}
    \end{array}\right\} \forall k, \forall j.
\end{equation}

\noindent In the above formulation, weights $w_{i,j}^{k-1}$ and biases $b_j^{k}$ are fixed (constant) parameters; the same holds for the objective function costs $c^k_j$ and  $d^k_j$. The ReLU function output for each unit is defined by \eqref{eq_relu_milp}, while \eqref{eq_bound_l0} and \eqref{eq_bound_k} define lower and upper bounds for the $x$ and $s$ variables: for the input layer ($k=0$), these bounds have physical meaning (same limits of the $Q_\theta$ inputs), while for $k\geq1$, these bounds are defined based on the fixed parameters $\theta$~\cite{ceccon2022omlt}. Notice that for the MIP formulation to be equivalent to the DNN, ReLU activation functions must be used, as explained in~\cite{fischettiJo2018}.

\subsubsection{Enforcing Constraints in Online Execution}

For an arbitrary state $s_t$, the optimal action $a_t$ can be obtained by solving the MIP in \eqref{eq_goal}--\eqref{eq_bound_k} derived from $Q_{\theta}$. In this case, as the decision variables are the actions $a_t$, the voltage magnitude constraints in \eqref{eq:voltage_boundary}, as well as the EESs SOC dynamics and the discharge/charge operation limits, in \eqref{eq_SOC_cons} and \eqref{eq_char_disc_cons}, respectively; are all added on top of the MIP formulation described by \eqref{eq_goal}--\eqref{eq_bound_k}. As a result, the optimal actions obtained by solving this MIP \textit{strictly} enforce all the actions and environment's operational constraints\footnote{A general mathematical proof of optimality and constraint enforcement for the proposed MIP-DRL framework is presented in our previous work in~\cite{shengren2023optimal}.}. This MIP problem can be represented as,
\begin{equation}\label{eq_online_execution}
\begin{aligned}
 \max_{\substack{a\in {\cal A}, x_j^k, s_j^k, z_j^k, \forall k}} \quad & \left\{ \eqref{eq_goal} \right\}\\
\textrm{s.t.} \quad & \eqref{eq_relu_milp}-\eqref{eq_bound_k},\eqref{eq_SOC_cons}, \eqref{eq_char_disc_cons}, \eqref{eq:voltage_boundary}. \\
\end{aligned}
\end{equation}

Algorithm~\ref{algorithm_online_execution} shows the step-by-step procedure used during the online execution of the proposed MIP-DRL framework. 

\begin{algorithm}[t]
	\caption{Online Execution for all the proposed MIP-DRL Algorithms}
	\label{algorithm_online_execution}
	Extract trained parameters ${\theta}$ from $Q_{\theta}$;\\
	Formulate the action-value function $Q_{\theta}$ as a MIP formulation according to \eqref{eq_goal}-\eqref{eq_bound_k}. Add on top all action space constraints i.e., \eqref{eq_SOC_cha}, \eqref{eq:voltage_boundary} and \eqref{eq_char_disc_cons}.\\
	Extract initial state $s_0$ based on real-time data;\\
	\For{$t=1$ \KwTo $T$}
	{For state $s_t$, get optimal action by solving \eqref{eq_online_execution} using commercial MIP solvers;
	}						
\end{algorithm}

\section{Simulation Results and Discussions}
\subsection{Simulation Setup}
\subsubsection{Environment Data and Framework Implementation}

\begin{figure}[t]
\centering
  \psfrag{A1}[lc][][0.8][0]{Load bus}
  \psfrag{A2}[lc][][0.8][0]{PV generation}
  \psfrag{A3}[lc][][0.8][0]{ESS}
\includegraphics[width=1.0\columnwidth]{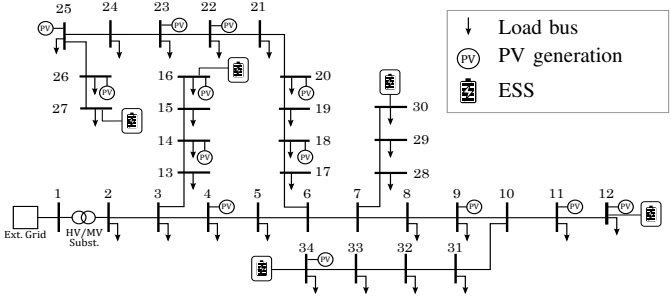}
\vspace{-2mm}
\caption{Modified IEEE-34 Node bus test system with distributed PV generation and EESs. The ESSs are placed at the end of each feeder to increase the number of voltage magnitude issues experienced.}
\vspace{-2mm}
\label{fig:ieee_system}
\end{figure}

To demonstrate the effectiveness of the proposed MIP-DRL framework, a modified 34-node IEEE test distribution network is used, as is shown in Fig.~\ref{fig:ieee_system}. ESSs are placed at nodes 27, 16, 30, 12, and 34 due to their higher chance of over- and undervoltage issues. The training data used corresponds to market day-ahead prices in the Netherlands, while consumption and PV generation measurements with a 15-min resolution are used. The voltage magnitude limits are defined as $\overline{V}=1.05$ and $\underline{V}=0.95$ p.u. PyTorch and OMLT (see ~\cite{ceccon2022omlt}) packages have been used to implement our MIP-DRL framework. Default settings shown in Table~\ref{tab:DRL_parameter} were used for all the implemented DRL algorithms during the value iteration procedure (in Sec.~\ref{value_iteration_procedure}). All implemented algorithms and the environment are openly available at~\cite{Shengren}.

\begin{table}[t]
\centering
\caption{Parameters for the DRL algorithms.}
\scalebox{0.7}{
\begin{tabular}{cccccccc} 
\hline
Algorithm & Batch size $|B|$ & Learning rate & Buffer size $R$ & $\gamma$ & Optimizer  \\ 
\hline
MIP-DDPG       & 512        & 6e-5          & 5e4         & 0.99                     & Adam~      \\
MIP-SAC        & 512        & 6e-5          & 5e4         & 0.99                    & Adam~    \\
MIP-TD3        & 512        & 6e-5          & 5e4         & 0.99                     & Adam~       \\
\hline
\end{tabular}}
\label{tab:DRL_parameter}
\end{table}

\subsubsection{Validation and Benchmarks for Comparison}
To assess the performance of the proposed MIP-DRL algorithms: MIP-DDPG, MIP-TD3, and MIP-SAC, their defined schedule is compared with their standard counterpart (DDPG, TD3). Two metrics are defined to do this: the operation cost, in EUR, and the number of voltage magnitude violations. Furthermore, we have also compared them with the optimal global solution obtained, considering a perfect forecast for the next 24 hours. This optimal solution is obtained by solving the NLP formulation in Sec.~\ref{sec_nlp_formulation}, implemented using Pyomo
. Notice that different from the optimal global solution, all the tested DRL algorithms (standards and their MIP counterpart) make decisions only using the current state information.


\vspace{-2mm}
\subsection{Performance on the Training Set}
\begin{figure}[t]
    \psfrag{Total Reward [-]}[][][1.0]{$\text{Total Reward [-]}$}
    \psfrag{A3}[][][0.6]{$\sum_{m \in {\cal N}} \left( P^D_{m,t}+P^{B}_{m,t}-P^{PV}_{m,t}\right)\text{[-]}$}
    \psfrag{A4}[][][0.6]{$\sum_{m\in\cal{B}} C_{m,t}(V_{m,t})\text{[-]}$}
    \centering
\includegraphics[width=0.85\columnwidth]{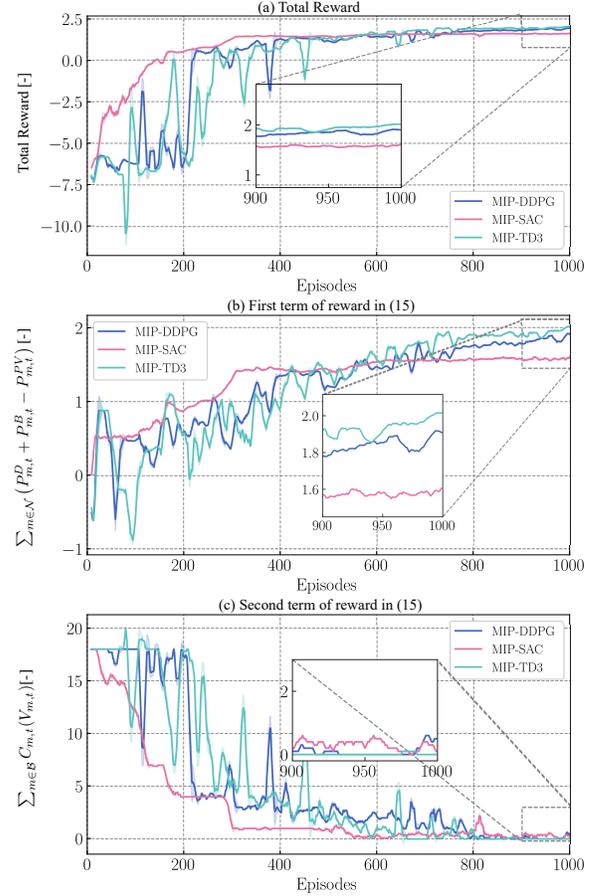}
    \caption{$(a)$ Average total reward as in \eqref{eq:new_penaly_reward}. $(b)$ Operational cost or first term of reward in \eqref{eq:new_penaly_reward}. $(c)$ Cumulative penalty for voltage magnitude violations or second term of reward in \eqref{eq:new_penaly_reward}, all during training.}
    \label{fig:training_data}
    \vspace{-3mm}
\end{figure}

Fig.~\ref{fig:training_data} displays the average total reward, operational cost, and the number of voltage magnitude violations during the training process for the developed MIP-DRL algorithms (MIP-TD3, MIP-DDPG, MIP-SAC). Results shown in Fig.~\ref{fig:training_data} are obtained as an average of over five algorithms executions. As shown in Fig.~\ref{fig:training_data}, the average total reward increases rapidly during the training, while simultaneously, the number of voltage magnitude violations decreases. This is a typical training trajectory of penalty-based DRL algorithms. At the beginning of the training process, the DNN's parameters are randomly initialized, and as a consequence, the actions defined cause a high number of voltage magnitude violations. Throughout the training, introducing a large magnitude penalty term in the reward definition in~\eqref{eq_reward} leads to updating the DNN's parameters, resulting in higher quality actions, primarily learning to reduce voltage magnitude violations, and, later on, improving the general performance. All three MIP-DRL algorithms converged at around 1000 episodes. The total reward of MIP-TD3 and MIP-DDPG converged at $2.01\mypm0.02$, and $1.94\mypm0.02$, respectively. Compared to MIP-TD3 and MIP-DDPG, that of MIP-SAC converged to a low value, at $1.57\mypm0.01$, indicating that MIP-SAC has a lower quality of actions. Notice that for MIP-DDPG and MIP-TD3, the operation cost significantly increases during the training process, while SAC does not improve after 400 episodes (Fig.~\ref{fig:training_data}). 

After the last training episode, the number of voltage magnitude violations of the MIP-TD3 algorithm is around 1. In contrast, a higher number of violations for the MIP-DDPG and MIP-SAC algorithms was observed at around 2. This result shows that DRL algorithms can effectively learn from interactions, reducing the number of voltage magnitude violations while minimizing the total operation cost by learning to dispatch the ESSs correctly. However, these trained policies \textit{cannot} strictly enforce voltage magnitude constraints. If such algorithms are used directly in real-time, they might lead to infeasible operation, causing voltage violations. Next, we will show how our proposed framework overcomes this during online deployment, enforcing all operational constraints even in unseen data.

\subsection{Constraint Enforcement Capabilities}

\begin{figure*}[!htbp]
    \centering
    \includegraphics[width=1.7\columnwidth]{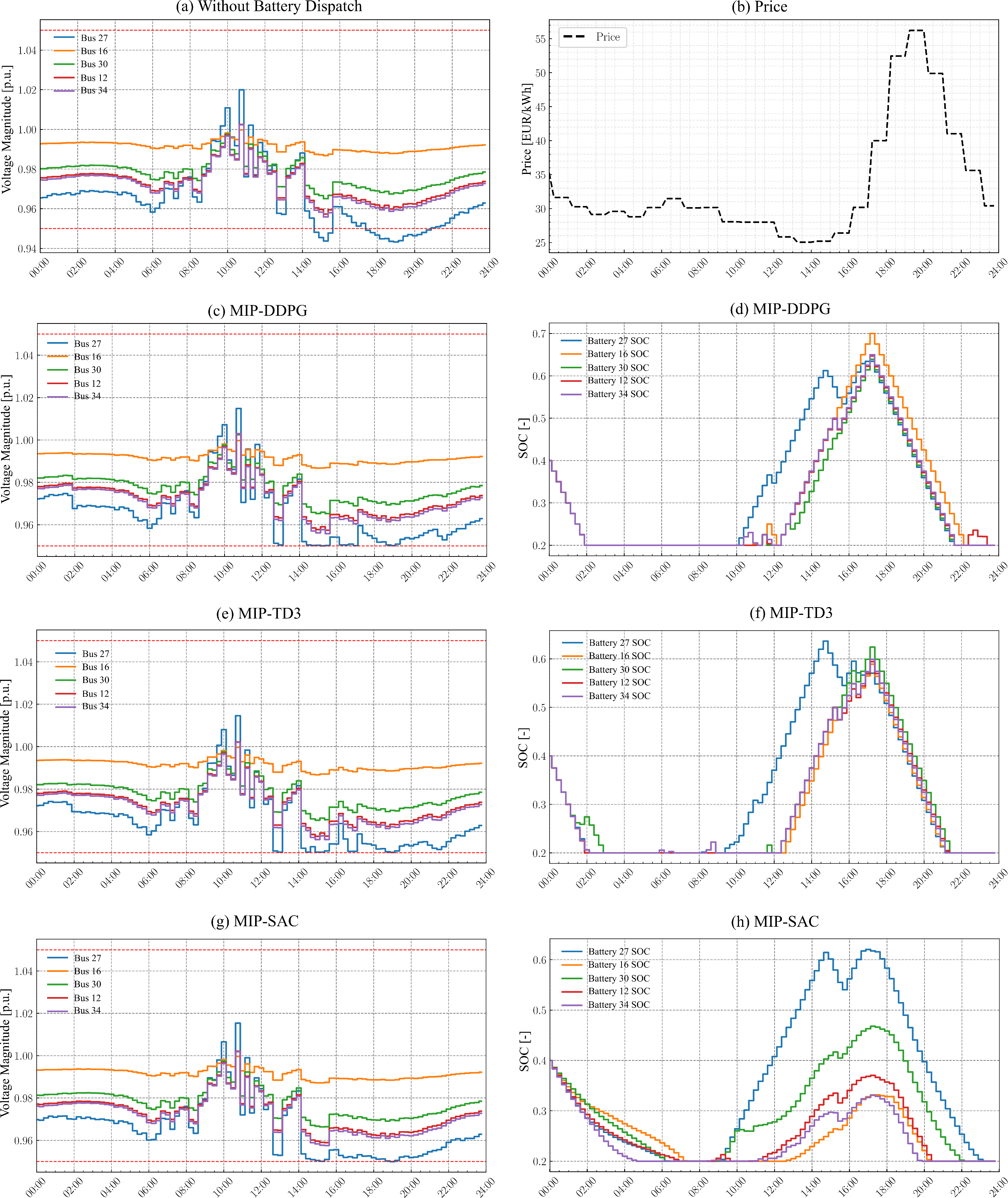}
    \caption{ (a): Voltage magnitude for nodes in which the ESSs are connected, disregarding their operation. (b): Price in \texteuro/kWh. Voltage magnitude ((c), (e) (g)) in which the ESSs are connected and SOC of ESSs ((d), (f), (h)), after executing the dispatch decisions provided by the MIP-DDPG, MIP-TD3, and MIP-SAC algorithms, respectively.}
    \label{fig:MIP_test}
\end{figure*}

\begin{figure}[!htbp]
    \centering
    \includegraphics[width=1.0\columnwidth]{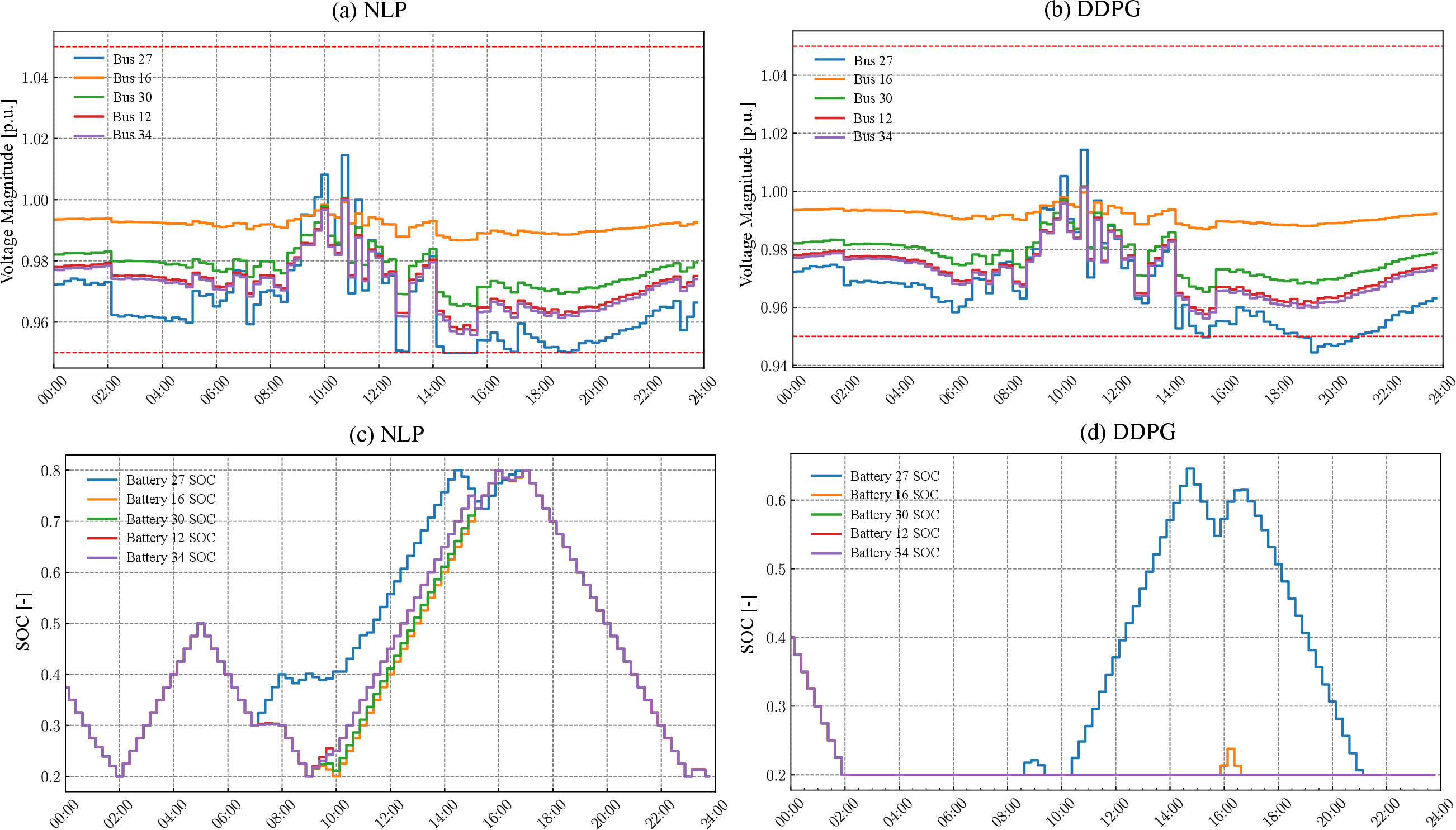}
    \caption{Voltage magnitude for nodes in which the ESSs are connected (a), (b) and SOC of ESSs (c), (d); after executing the dispatch decisions provided by NLP mathematical formulation with perfect forecast (optimal solution) and the standard DDPG algorithm, respectively.}
    \label{fig:NLP_DRL_test}
\end{figure}


Fig.~\ref{fig:MIP_test} displays the voltage magnitude of the nodes in which the ESSs are connected and the ESSs' SOC during a typical day in the test dataset. The results showed in Fig.~\ref{fig:MIP_test} are obtained after using the dispatch decisions provided by the MIP-DDPG, MIP-TD3 and MIP-SAC algorithms. Fig.~\ref{fig:MIP_test}~$(a)$ shows the voltage magnitude of the nodes in which the ESSs are connected, but in this case, disregarding their operation (i.e., ESSs are neither charging nor discharging), while Fig.~\ref{fig:MIP_test}~$(b)$ shows the day-ahead electricity price of that test day. As can be seen in Fig.~\ref{fig:MIP_test}~$(a)$, if the ESSs's operation is disregarded, the voltage magnitude at node 26 faces undervoltage problems between 14:00-16:00 and 18:00-20:30. Thus, a proper dispatch of the available ESSs must enforce that such voltage magnitude constraints are met. This is the case when executing the dispatch decisions provided by the developed algorithms, as shown in Fig.~\ref{fig:MIP_test}~$(c)$ for the MIP-DDPG algorithm, in Fig.~\ref{fig:MIP_test}~$(e)$ for the MIP-TD3 algorithm and in Fig.~\ref{fig:MIP_test}~$g$ for the MIP-SAC algorithm. As all the developed MIP-DRL algorithms dispatch the ESS connected at node 26 in discharging mode during the above-mentioned time periods, all undervoltage issues are solved. In terms of dispatch decisions, and as seen in Fig.~\ref{fig:MIP_test} $(d)$, $(f)$ and $(h)$, all the developed MIP-DRL algorithms learn to first discharge all ESSs to the minimum SOC during the period between 00:00 and 06:00. Then, all ESSs are dispatched in charging mode during the period between 10:00 and 17:00, when the price is low, to then operate in discharging mode during the period between 16:00-22:00. This operational schedule during the peak consumption period reduces the amount of power the external grid provides while simultaneously solving the undervoltage issues. Compared to the MIP-DDPG algorithm, the MIP-TD3 and MIP-SAC algorithms provide more conservative dispatch decisions, leading to higher operational costs. The operational cost resulting from the dispatch defined by the MIP-DDPG algorithm is 13.87~$k$\texteuro, 3.1\% lower and 7.5\% lower, than the dispatch defined by MIP-TD3 and MIP-SAC algorithms, respectively. These results are due to each MIP-DRL algorithms different policy exploration strategies.




Fig.~\ref{fig:NLP_DRL_test}~$(a)$ and~$(c)$ display the voltage magnitude of the nodes in which the ESSs are connected and the ESSs' SOC, respectively, after executing the dispatch decisions provided by NLP mathematical formulation with perfect forecast (optimal solution). Compared with the decisions defined by the NLP formulation considering perfect forecast, the dispatch decisions provided by the MIP-DRL algorithms in Fig.~\ref{fig:MIP_test} show a similar operational pattern. Notice in Fig.~\ref{fig:MIP_test}~$(d)$,~$f$ and~$h$ that when the electricity price is low, the proposed MIP-DRL algorithms dispatch the ESSs in charging mode, while in discharge mode, when the electricity price is high. A similar dispatch decision is observed in the optimal global solution, as shown in Fig.~\ref{fig:NLP_DRL_test}~$a$. Notice also that, when compared with the optimal solution, the proposed MIP-DRL algorithms dispatched conservative solutions, as these algorithms avoid charging all ESSs to the maximum SOC when the electricity price is low. This can be considered a sub-optimal decision. In this case, the operational cost resulting from the dispatch defined by the MIP-DDPG, MIP-TD3 and MIP-SAC are 9.5\%, 12.9\%, and 18.4\% higher, respectively, than the dispatch defined by the NLP formulation. The difference in this dispatch decision can be due to the estimated action-value function, which might not be good enough to represent the true action-value function. As the proposed MIP-DRL algorithms choose actions that maximize its $Q$-value estimation, the largest $Q$-value might not represent the best action for this specific state-action pair. Nevertheless, even in executing a sub-optimal decision, the proposed MIP-DRL algorithms enforce all voltage magnitude constraints, guaranteeing operational feasibility. Additionally, although differences in the dispatch decisions made by the proposed MIP-DRL algorithms and the optimal solution can be observed, it is important to highlight that the optimal solution is obtained considering the perfect forecast of the future generation and demand consumption for the next 24 hours, while the proposed MIP-DRL algorithm provides dispatch decisions in a quarter-hourly basis, without knowledge of the future values of the stochastic variables. 



Finally, if the dispatch decisions defined by the developed MIP-DRL algorithms are compared with the ones provided by the standard DDPG and TD3 algorithms, lower operational costs are observed while all voltage magnitude constraints are strictly enforced. This can be seen in Fig.~\ref{fig:NLP_DRL_test}~$(b)$ for the DDPG algorithm (results for the TD3 algorithm are not presented due to space limitation). In this case, the cost of the dispatch decisions defined by the standard DDPG and TD3 algorithms are 22.3\% and 14.3\% higher, respectively, than the ones defined by their MIP counterpart. These higher costs are due to the failure of the standard DRL algorithms to dispatch all the ESSs. In general, the trained policy of these standard DRL algorithms learns to define an action that maximizes the Q-value under the current state, as shown in~\eqref{eq_DDPG_policy} and~\eqref{eq_TD3_policy}. This maximization is achieved by approximation and gradient ascent, which lacks the optimality guarantee, ultimately degenerating the quality of defined actions. Moreover, and as observed, these defined actions have no feasibility guarantee, leading to infeasible actions. In contrast, the proposed MIP-DRL algorithms deliver better dispatch decisions (in terms of the total operational cost) even if they have the same training process as the proposed MIP-DRL algorithms. The main reason for such improved performance is that the MIP-DRL algorithms directly solve the maximization of the Q-value problem. However, this is only possible if the DNN is transformed as a MIP formulation, as explained in Sec.\ref{sec:explain_DNN}.

\subsection{Error Assessment and Computational Performance on the Test Set}
Table~\ref{tab_compare_time} presents the average total error (with respect to the solution obtained by solving the NLP formulation with perfect forecast) for the operational cost, the average number of voltage magnitude violations, and total average computational time of the proposed MIP-DRL algorithms as well as benchmark DRL algorithms, over a period of 30 (unseen) test days. As can be seen in Table~\ref{tab_compare_time}, the proposed MIP-TD3, MIP-DDPG, and MIP-SAC algorithms can strictly enforce the voltage constraints. Among all these MIP- DRL algorithms, MIP-DDPG has the lowest average error, 10.4\%. In contrast, their standard counterparts, such as DDPG, showed poor performance reaching an error of 34.3\%, and violating the voltage magnitude constraint in around 45 time steps. As expected, the computational time required to execute the proposed MIP-DRL algorithms is higher than standard DRL algorithms. This increase in the computational time results from the MIP formulation needed to be solved to enforce all the operational constraints (see \eqref{eq_online_execution}). Nevertheless, for this case, the proposed MIP-DRL algorithms can still be used for real-time operation as it only requires less than 60 seconds for one day (96 time-steps) execution.

\begin{table}[t]
\centering
\caption{Performance comparison of different DRL algorithms in an unseen test set of 30 days.}
\label{tab_compare_time}
\scalebox{0.9}{
\begin{tabular}{ccccl}
\hline
\multicolumn{1}{l}{Algorithms} & \begin{tabular}[c]{@{}c@{}}Operation \\ Cost Error {[}\%{]}\end{tabular} & \begin{tabular}[c]{@{}c@{}}Voltage Magn.\\ Violations {[}-{]}\end{tabular} & \begin{tabular}[c]{@{}c@{}}Comp.\\ Time $[s]$\end{tabular} \\ \hline

MIP-TD3 & $13.2\mypm0.5\%$ & 0& 57$\mypm$6.7           &  \\
MIP-DDPG             & \textbf{10.4$\mypm$0.7\%}          & 0       & 43$\mypm$5.1          &  \\
MIP-SAC              & 19.3$\mypm$1.5\%          & 0       & 57$\mypm$6.3 &  \\
TD3              & 28.5$\mypm$0.4\%          & 33$\mypm$2       & 16$\mypm$0.1          &  \\ 
DDPG             & 34.3$\mypm$0.7\%          & 45$\mypm$11       & 16$\mypm$0.1          &  \\ 
\hline
\end{tabular}}
\vspace{-2mm}
\end{table}

\subsection{Discussion}

We have successfully combined deep learning and optimization theory to bring constraint enforcement to DRL algorithms. By using the trained Q-network as the surrogate function of the optimal operational decisions, we have guaranteed the optimality of the action from the Q-network through the formulated MIP. Moreover, by integrating the voltage constraints into the formulated MIP, the feasibility of the action is enforced. However, the performance of MIP-DRL algorithms is determined by the approximation quality of the Q-network, obtained after the training process is performed. During this training process, the Q-iteration faces the exploration vs. exploitation dilemma, which can impact the approximation quality. For instance, the MIP-DDPG algorithm outperforms the MIP-TD3 algorithm, while the MIP-SAC algorithm performs poorly in the framework. This discrepancy may be caused by the divergence between the exploration policies leading to different exploration efficiencies and Q-networks update rules. The conservative performance of the MIP-SAC algorithm might be caused by the soft Q updating rule, which introduces more assumptions, impacting the estimation for accurate approximation.

\section{Conclusion}
This paper proposed a DRL framework, namely MIP-DRL, to define high-quality dispatch decisions (in terms of the total operational cost) for battery storage systems within a distribution network, while ensuring their technical feasibility (related to enforcing voltage magnitude constraints). The proposed framework consists of a Q-iteration and deployment procedure. During the Q-iteration procedure, a deep neural network (DNN) is trained to represent the accurate state-action value function. Then, during the deployment procedure, this Q-function DNN is transformed into a MIP formulation that can be solved by commercial solvers. Results showed that the dispatch decisions defined by the proposed MIP-DRL algorithms can ensure zero voltage magnitude violations while standard DRL algorithms fail to meet such constraints in unseen data. Additionally, the proposed MIP-DRL algorithms showed low errors, near to 10.4\% (for the MIP-DDPG algorithm), when compared with the optimal solution obtained with a perfect forecast of the stochastic variables. 

\bibliographystyle{IEEEtran}
\bibliography{MIP_DRL_Framework_Paper} 
\vspace{-2mm}


\end{document}